\documentclass[journal]{IEEEtran}
\input epsf
\usepackage{graphicx}

\begin{document}

\title{An octave bandwidth frequency independent dipole antenna}

\author{AGARAM~RAGHUNATHAN, N.~UDAYA~SHANKAR, RAVI~SUBRAHMANYAN 
\thanks{Manuscript received Jan, 2013.} 
\thanks{The author is with the Raman Research Institute, C.V.Raman Avenue, Sadashivanagar, Bangalore-560085, 
India (e-mail: raghu@rri.res.in).}}

\markboth{ }
{Shell \MakeLowercase{\textit{et al.}}: 
An octave bandwidth dipole antenna}

\maketitle

\begin{abstract}
Precision measurements of the spectrum of the cosmic radio background require
frequency independent antennas of small electrical dimensions.  We describe the
design of a wide-band fat-dipole antenna with a sinusoidal profile having a 
frequency independent performance over the octave band 87.5 to 175~MHz. The input
return loss exceeds 15~dB and the radiation power pattern is frequency invariant and
close to cosine square over the octave bandwidth. The structure has been optimized
using electromagnetic modeling, and the design has been validated by constructing a
prototype.\end{abstract}


\section{Introduction}

\PARstart{A}{} simple half-wave dipole antenna works on the principle of resonance;
therefore, it inherently has a narrow bandwidth and its radiation pattern and
impedance change rapidly away from the resonant frequency. In general, any antenna
structure whose thickness varies smoothly over its length would radiate energy
efficiently over a wide frequency range.  However, its impedance and radiation
characteristics would be independent of frequency only if its surface current decays
linearly away from the feed point in a frequency dependent fashion [1]. 
Several such antenna structures that exhibit frequency independent performance have
been discussed  in the literature [2]. 

The scientific motivation of designing a frequency independent antenna is to measure the brightness spectrum of the cosmic radio background and faint features in the spectrum that arise from redshifted atomic hydrogen 21cm emission. The brightness spectrum is expected to have structures over a range of scales. If the antenna used to measure has frequency dependent radiation pattern, it results in spurious spectral response similar the red shifted 21cm emission. To avoid this problem, a frequency independent antenna is required [3].

With the aim of deriving a frequency independent antenna for the  measurement of
cosmic radio background in the frequency range 87.5--175~MHz, we carried out electromagnetic simulations of a few types of
smoothly profiled fat-dipole structures and examined their performance for frequency
independent behaviour.  We used WIPL-D electromagnetic (EM) simulation software for
modeling. Our study showed that a dipole with a sinusoidal profile exhibits  good
frequency independent characteristics over the octave band  87.5 to 175~MHz.  A
prototype of the fat-dipole operating in this  frequency range  has been built and
tested for its frequency independent performance.  Section II of this paper describes our design
objectives motivated by observational cosmology. Our investigations of a few types
of smoothly profiled fat-dipole structures, especially the sinusoidal profile in detail,
are described in Section III. Section IV contains the details of fabrication of
the prototype dipole. The antenna performance measurements  are
described in Sections V to VIII and our work is summarized in section IX.  

\section{Design objectives}
The motivation for the antenna design is in its application for a wide band spectral
measurement of the cosmic radio background; specifically for the detection of
features arising from events in the epoch of reionization [4],~[5]. Cosmological expansion of the Universe shifts the 1420.4
MHz spin flip transition of neutral atomic hydrogen gas, which is the most abundant
element in early times and in the epoch of reionization, by factors between 8--15
into the octave band 87.5--175~MHz. The cosmological signals are 
predicted to be at most 20--30~mK and hence about $10^4$ times
weaker than the foreground emission from celestial radio sources. Nevertheless, they
may be distinguished by their frequency structure that is in contrast with the
relatively smoother foregrounds [6], which arise from 
impulsive thermal brehmsstrahlung
and synchrotron processes that are inherently broad spectrum emissions. In our application, the designed dipole will be used to estimate the
cosmological signal averaged over a wide field of view in the sky. This
will be carried out by measuring the total power from the sky using the
designed dipole and subtracting from it a model for the foreground
emission of the sky and the receiver noise derived from the same set of observations [3].

Two characteristics of an antenna element relevant for detecting weak spectral signatures
in the sky signal are
i)~Radiation  pattern and ii) Return loss. Their dependence on  frequency 
plays a major role in limiting the detectability of such signals. 

To arrive at the design specifications of the dipole, we model the sky temperature distribution using the 408~MHz sky 
survey of Haslam et al. [7].  We also assume a spectral index distribution that is
given by the brightness ratio between this image and that of the 150-MHz 
all-sky map synthesized by Landecker et al. [8].  When averaged over a significant
sky area, the variation in log temperature versus log frequency is well fit by a third order polynomial.
However, the brightness variations over the sky will naturally result in undesirable
spectral structure and enhanced residuals to the above fit 
if the measuring antenna has a radiation pattern that varies with frequency. 
We have computed the increase in residuals to the above fit assuming frequency dependence in the antenna beam
pattern and find that residuals would be less than 5~mK (a small fraction of the signal expected) 
if the half-power beamwidth varies by less than $\pm 2.5\%$
over the  frequency range 87.5 to 175 MHz.  This defines the tolerance for 
frequency dependence of the radiation power pattern of the fat dipole antenna.
Since in general an electrically small radiator exhibits a frequency independent radiation  pattern,
we prefer to have an antenna which is electrically small in dimensions.

An antenna with relatively constant impedance is
preferred, so that the antenna gain losses as well as internal reflections of
the receiver noise are low and vary smoothly with frequency. 
For efficient performance, a return loss of 10~dB or more  is preferred over the entire frequency band. 
However, even with this return loss, a 0.1~dB 
variation in it results in a passband ripple of about 0.5~K,~for an antenna temperature of 200~K.
A carefully crafted bandpass calibration methodology is required to eliminate this problem; nevertheless,
a low and relatively uniform return loss without  any inflection points in the band of interest 
is desirable to relax the calibration requirements. { Several broadband antennas are available off-the-shelf for Eg. Quadridged horn antenna from ETS Lindgren(www.ets-lindgren.com),~magnetic mount antenna from Antenna Products (www.antennaproducts.com) and omni directional wide band antenna from AntennaSys Inc.(www.antennasys.com) which have bandwidth more than an octave. However, they do not seem to have frequency independent beam characteristics and smoothly varying impedance match over their operating band. These two aspects have been achieved in the work presented in this paper.

 \section{Investigation of smoothly profiled fat dipole structures }
Frequency independent antenna characteristics are exhibited by radiating
structures whose shapes are specified only in terms of angles. Additionally,
biconical antennas that are infinitely long and specified only in terms of the included cone angle 
also exhibit frequency independence characteristics[9].  However, because of their unmanageable dimensions, they are not practically realizable. Meeting the
frequency independent requirement in a finite structure requires that the
current attenuates along the structure and be negligible at the point of
truncation. The profile of the antenna structure is expected to play an important role in attenuating the surface
current. In addition, it also controls the radiation reactance of the antenna. A profile
that gives minimum radiation reactance is always preferred. 

 Several authors have investigated the broad band performance of planar fat dipole structures having different profiles [10-12]. Those structures are found to have non-uniform radiation patterns with large dispersion in their half
power beamwidths and higher order impedance responses, which are unacceptable for the application on hand. In view of this,
we investigated the frequency-independent performance of non-planar dipoles having five different smooth profiles :
i) exponential ii) conical with multi-conic sections  iii) cubic spline 
iv) sine-square and v) sinusoidal. The investigation was carried out through simulations 
using WIPL-D - an electromagnetic modeling and
simulation software (www.wipl-d.com) used extensively for accurate frequency-domain analysis of metallic
structures. 

In our simulation,~the first three profiles for a two point excitation showed a variation
in the radiation reactance in
the range $-100$ to +100~ohms  and the fourth one showed a variation in the range of 0 to 35 ohms
in the frequency range 87.5 to 175~MHz.
However,  with the sinusoidal profile, this variation was limited to the range  -10 to +10
ohms. This led us to simulate and explore the sinusoidal profile in greater
detail.

Each arm of the fat dipole simulated,  has basically three sections : 
i) input conic section ii) profiled middle section 
and iii) terminating conic section. A finite gap separates the two arms and provides space
for making electrical connections.

 The input conic section varies  the thickness of the dipole smoothly  at the
feeding point. This gradual variation helps in maintaining  input impedance of the dipole constant over a wide range of frequencies.

 The profiled middle section controls i)~the generation of 
the reactive component of the 
antenna impedance and ii) the dependency of the radiation pattern  on frequency. 
The profile used in our design is described by the equation :

$$x= r_{0} + Asin^{\alpha}(kz)         \eqno (1) $$
where

\indent $r_{0}$ is the radial shift given to the profile while making the dipole thick \newline
\indent $A$ is the amplitude of the profile defined in terms of operating wavelength, \newline
\indent $\alpha$ is the index of the sinusoidal function,  \newline
\indent $k$ = 2$\pi$/$\lambda$ and \newline
\indent $\lambda$  is the design wavelength. Fig.~$\ref{ant_3D}$ shows a 3-D view of the structure investigated, with its various
parameters indicated.
~~A short dipole exhibits frequency independent characteristics and since we aim to design a profiled dipole that is
short over an octave bandwidth, the design frequency 
is biased towards the higher frequency end of the band.

The conic section 
at the termination minimizes abrupt reflection of surface current.  Any  abrupt reflection is expected to 
narrow the operating bandwidth; therefore,  the output conic section is profiled smoothly like the input conic section.

 The body of revolution of the dipole has been approximated by
four flat aluminium sheets each bent along a { sinusoidal} profile. This simplification was  adopted to 
ease the manufacturing process  of the dipole and make disassembly and transportation to remote observing sites 
easy. The entire simulation study was carried out for this simplified structure.

The simulation was employed mainly  to understand the influence of i) various physical
parameters of the proposed dipole structure, and ii) the design wavelength, on the input return loss and radiation pattern 
of the antenna.  
We examined physical parameters like a) the gap between the two arms of the dipole
structure, b) the dimension of the conic sections at the feeding point and termination
and c) functional form of the profile and its amplitude. Each parameter
was varied over a finite range based on constraints posed by physical
realizability.

\begin{figure}

\includegraphics[width=3.2in]{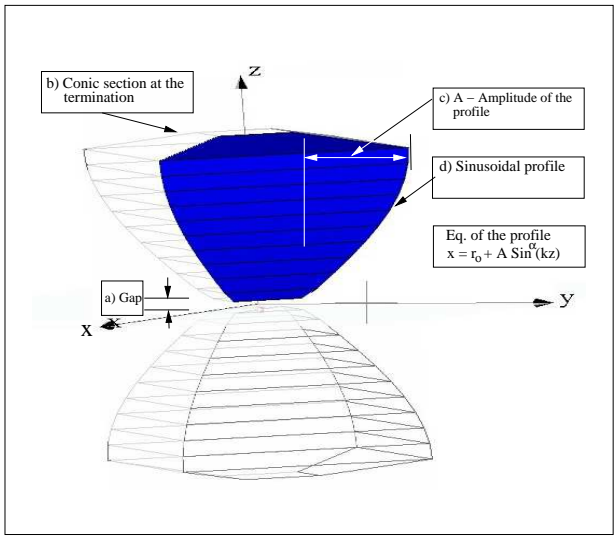}
\caption{A three dimensional view of the fat dipole  structure with sinusoidal profile; 
the various parameters defining the structure are indicated. }
\label{ant_3D}
\end{figure}

\begin{figure}
\includegraphics[width=3.2in]{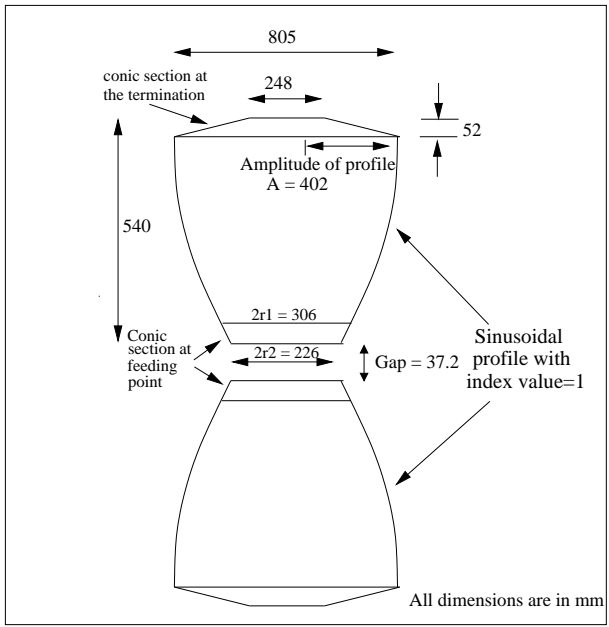}
\caption{Schematic indicating the physical dimensions of the arms of the fat dipole antenna}
\label{ant_sch}
\end{figure}

In our simulation, optimization of the structure was not carried out using the entire parameter space available, to 
obtain a global minimum. Instead, a technique similar to the gradient descent was adopted to optimize each of the system parameters.
In this method, each parameter was varied from its initial value assigned over a range decided mainly by 
the fabrication considerations, in the direction of
the negative gradient of the vector  till a local minimum was reached. The gradient vector represents the 
change in the antenna performance as a function of  physical parameters.

The antenna performance was quantified by estimating
the variation of 3-dB beamwidth as a function of frequency and visually inspecting the return loss for its
smooth and flat response. Fig.~$\ref{ant_sch}$ shows the schematic diagram of the fat dipole with optimized structural 
parameters.

 Normally a resonating antenna exhibits frequency independent characteristics when it is used as a short dipole. Therefore, if we ensure that the electrical length of the antenna is smaller than half wave length at the highest frequency of operation, then the antenna would act as a short dipole over most of the operating band. However, a short dipole does not have an easily matchable impedance characteristic even over an octave bandwidth.

 The  wavelength for defining the electrical dimension of the structure was varied in simulation over a wide range in the operating band, covering the entire band of interest trying to obtain both frequency independent radiation characteristics and lower order return loss response. After simulation, 2027~mm was
 obtained as an optimum  design wavelength ($\lambda_{o}$)
for which the dipole exhibited a minimum deviation  ( $<$~2 \%) of 3-dB beamwidth across the entire frequency band and greater than 10 dB return loss.  
2027~mm corresponds to a frequency of 148 MHz. As expected, this is much closer to 175 MHz  which is the maximum frequency of operation.

In order to study the behaviour of the dipole,~each parameter was varied about its 
optimal value and at each step the system response was carefully monitored. 

The gap between the two arms of the dipole structure was varied
between 20 mm to 56 mm. It was found that it had
a major influence on the return loss of the antenna.  A gap lesser
than 30~mm resulted in a return loss exceeding 15~dB only at lower frequencies while a gap  more than 40~mm improved return loss only at higher frequencies.
However, a gap of 37.2~mm resulted in a return loss  $\geq15$~dB throughout the frequency range. 
Fig.~$\ref{gap_effect}$ shows the variation 
of the input return loss  over the frequency range 87.5--175~MHz for a range of  gaps between the two  arms of the dipole structure.
In this plot other parameters
of the antenna structure were maintained at their optimal values obtained in the design study (see Fig.~$\ref{ant_sch}$). 
The gap  did not appear to have any significant effect on the radiation patterns at different frequencies.

\begin{figure}
\includegraphics[width=6.2cm,angle=-90]{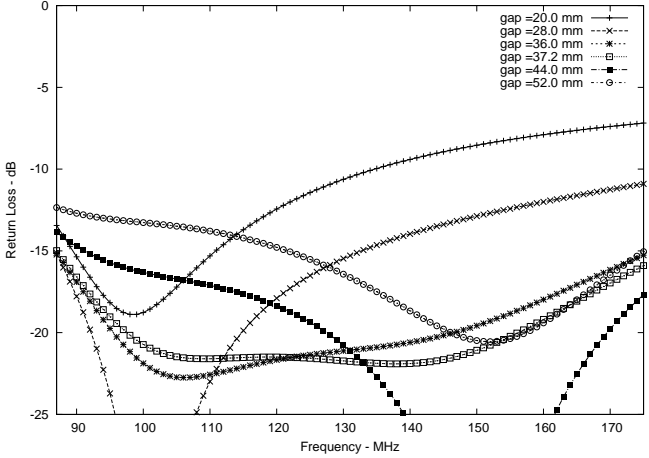}
\caption{Plot showing the effect of gap between the two halves of the dipole on the 
input return loss of the antenna. A gap of 37.2~mm between the dipole structures results in a  return loss $\geq15$~dB throughout
the frequency range.}
\label{gap_effect}
\end{figure}

The dimension of the  truncated conic section  at the feeding point also influences the
return loss substantially. The radius of the conic section at the feeding point was varied from 60 mm to 150 mm. 
It was noticed that the return loss had a subtantial variation in the frequency range 87.5 to 175~MHz both at lower and
higher values of the radius. For  a radius (r2) of 113~mm at the feeding point
and 153~mm (r1) at the start of the flare, return loss observed exceeded 15 dB over the entire band and had
a uniform response with frequency.

The index of the sinusoidal function used to form the profile of the antenna
structure was varied in the range 0.5--1.5. Lower index value improved the return loss at lower frequencies but degraded it at higher frequencies
whereas
higher index  value improved return loss at higher frequencies but worsened it at lower frequencies.
An index value of one was observed to be  optimum for a  return loss $\geq 15 dB$ over the entire frequency band. 
The antenna  response for various index values are shown in Fig.~$\ref{index_effect}$. The deviation of 3-dB beam width  was found to be more than
2~$\%$ at both lower and higher index values. So to satisfy both the requirements of good return loss and also
uniformity in beam pattern, an index value of 1 was selected for the sinusoidal profile.

\begin{figure}
\includegraphics[width=6.2cm,angle=-90]{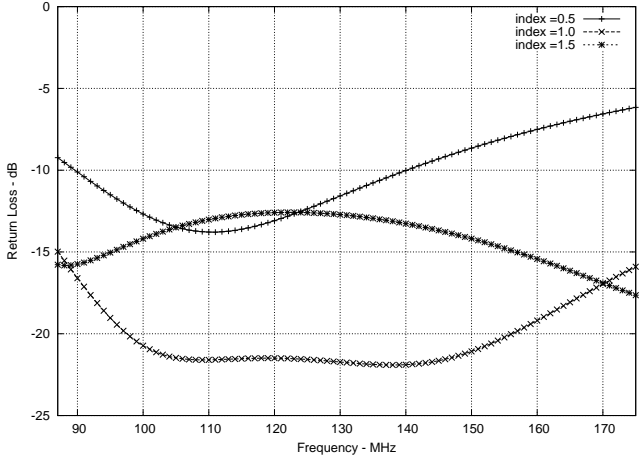}
\caption{Plot showing the effect of change in index of the profile of the dipole structure on
the input return loss of the antenna.}
\label{index_effect}
\end{figure}

The amplitude of the profile ($A$) is another parameter that has substantial influence on  
both the input return loss and 3-dB beamwidth at various frequencies.  The initial value assigned 
to it was half the electrical length of each arm ($\frac{\lambda}{8}$) at the designed optimum wavelenth. When the magnitude of the amplitude was varied 
around this value, it was found that the required characteristics like  good return loss ($\geq 15 dB$) and frequency independent behaviour 
of the radiation pattern could be achieved only over a narrow range of amplitude values.~Both at lower and higher values, return loss 
exhibited sharp resonances resulting in  large variation of return loss over a narrow range of frequencies. In addition,
the variation of 3-dB  beam width was 
also more than 5~$\%$ across the entire band. 
The optimum value determined for the amplitude of the profile was 1.2 $\frac{\lambda}{8} $; for this choice 
the return loss showed uniform response with frequency
and the deviation in 3-dB beamwidth was less than 5$\%$.

The dimension of the conic section at the termination did not have any
noticeable influence on either the radiation characteristics or the input return loss. This
indicates that the surface currents near the end points are negligible over the entire frequency
range of operation and  hence does not affect the frequency-independent performance. In a similar
way, the conic section at the feeding point also did not have any influence on the radiation pattern.

The simulation of the entire antenna structure was carried out without having any ground reflector below. 
The presence of a reflector at a finite distance makes the response of the antenna highly frequency dependent
and, therefore, the antenna is intended for use above an absorber plane.

\section{Fabrication of a prototype of the fat dipole antenna with a Sinusoidal profile}

 As a first step in the process of fabrication, the  length  of the lateral surface (L) and  the thickness (W) of the dipole structure are determined. An aluminium sheet measuring L meter long and W meter wide is taken and kept on a graph sheet. Assuming that the graph sheet  represents the XoZ plane, the lower edge of the sheet is  kept aligned along  the x-axis. The sheet is displaced from the x-axis in the  +ve z direction by a distance equal to half the  centre gap of the dipole structure. 

At every point on the profile of the structure, the width of the sheet is calculated from the relationship : Width = 2*x*cos(45), where x represents the x-co-ordinate of the point (Refer Eq. 1) considered on the profile. The angle 45 deg. used in the above relationship assumes that the dipole structure has square cross section.~The widths obtained  are transferred to the sheet at their respective distances from the bottom edge of the sheet. In a similar way, the widths at the feeding and terminating sections  are  also determined and transferred to the sheet. After completing this process, the sheet is cut along the transferred marks to obtain a shape as shown in Fig.~$\ref{ant_sheet}$. 

To bend the sheet, the distance of every point on the profile from the z-axis (antenna axis) is found out using the relationship : Distance from antenna axis = x*cos(45). The sheet is bent at every point marked on it, maintaining the distance determined above from the antenna axis (Refer Fig.~$\ref{ant_edge}$). This process is repeated for eight aluminium sheets.~Once they are cut and bent, they are fastened along their edges using bolts and L-brackets to get the desired shape as shown in 
Fig.~$\ref{ant_photo}$. The two halves of the dipole were
held rigidly together using non-electrically conducting bakelite spacers. This was done to minimize mechanical stress on 
the feed terminals while making electrical connections between the antenna and the balun. The entire antenna  was
mounted on light weight styrofoam to ensure that the antenna performance is not affected by the supporting structure. 

Tags mounted at the center of the square blocks at the inner ends of the two dipole arms serve as feed points. These tags are normally connected to the inputs of a transformer type balun for converting the balanced output of the antenna to unbalanced output. 

 However, imperfect  electrical characteristics of transformer type  balun prevent the true impedance of the 
antenna to appear at its output. 
This problem was overcome by  using a choke balun formed out of a RG-174 co-axial cable with ferrite beads.
Fig.~$\ref{ant_connect}$ shows the photograph of the feeding section in which the antenna output is connected to the choke balun.
Measurements made to estimate the resistive loss  due to the ferrite beads revealed that
 the attenuation was less than 1 dB.  The antenna signal is expected to get attenuated by this much before reaching the receiver input 
terminals.

Antenna measurements were made on the prototype to characterize the radiation and
impedance properties of the antenna; these are described below.   

\begin{figure}
\includegraphics[width=3.2in, height=3.2in]{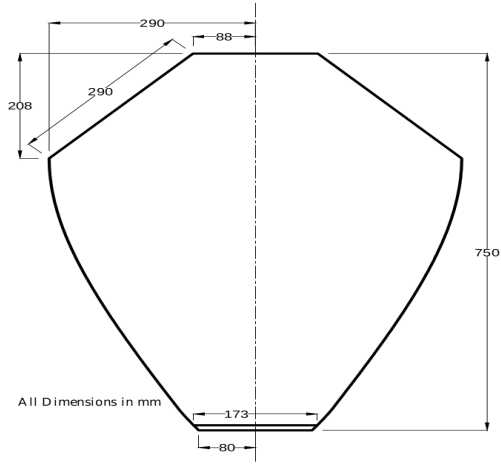}
\caption{Schematic indicating the dimensions of sheet forming the dipole structure before bending}.
\label{ant_sheet}
\end{figure}

\begin{figure}
\includegraphics[width=3.2in, height=3.2in]{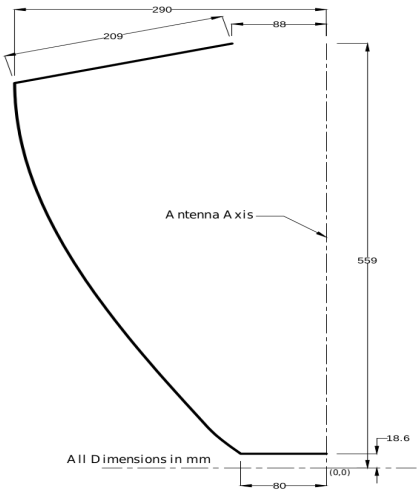}
\caption{Schematic indicating the profile along which the sheet is to be bent to form the dipole.}
\label{ant_edge}
\end{figure}

\begin{figure}
\includegraphics[width=3.2in]{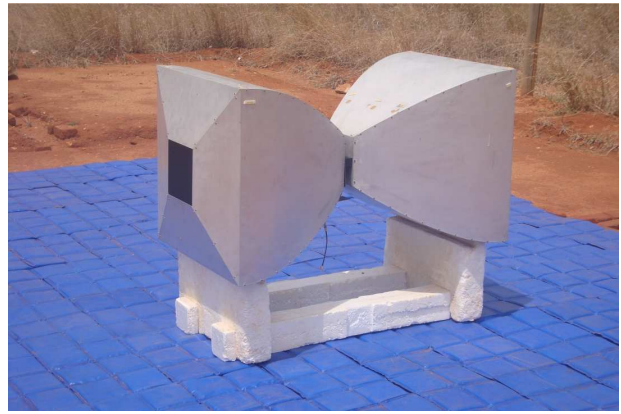}
\caption{Front view of the prototype dipole antenna with a sinusoidal profile.}
\label{ant_photo}
\end{figure}

\begin{figure}
\includegraphics[width=3.2in]{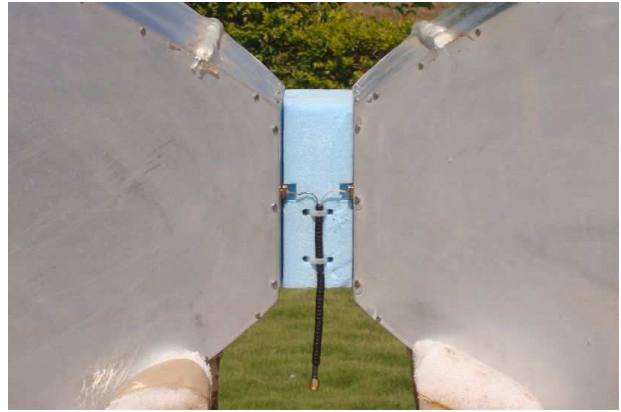}
\caption{Photograph illustrating the electrical connection made at the feeding point of the antenna.}
\label{ant_connect}
\end{figure}

\section{Measurement of Return loss }

Return loss is an important performance indicator of an antenna. It indicates the loss while coupling the sky signal from free space via the antenna to its feed terminals.   It is primarily 
due to impedance mismatch between i)~the free space and the antenna as well as
 ii)~the antenna and the transmission line connected to its terminals .
One of the main objectives in wide-band antenna design is to 
minimize this impedance mismatch and consequent signal loss across the band so that sky signal  coupling efficiency of the antenna is maximized.

The return loss was measured using Agilent Field Fox RF analyser N9912A.  The antenna under test
was placed over absorber ferrite tiles in order to minimize ground reflection and the measurement was done in an open field.
The ferrite tiles were sourced from Panashield Inc.(Model No. SFA 6.3) and were selected to be good absorbers in the frequency range of interest: 
they reflect only about 0.3$\% $ of the incident radiation in the frequency range 87.5 to 175~MHz.

\begin{figure}
{\includegraphics[width=6.2cm,angle=-90]{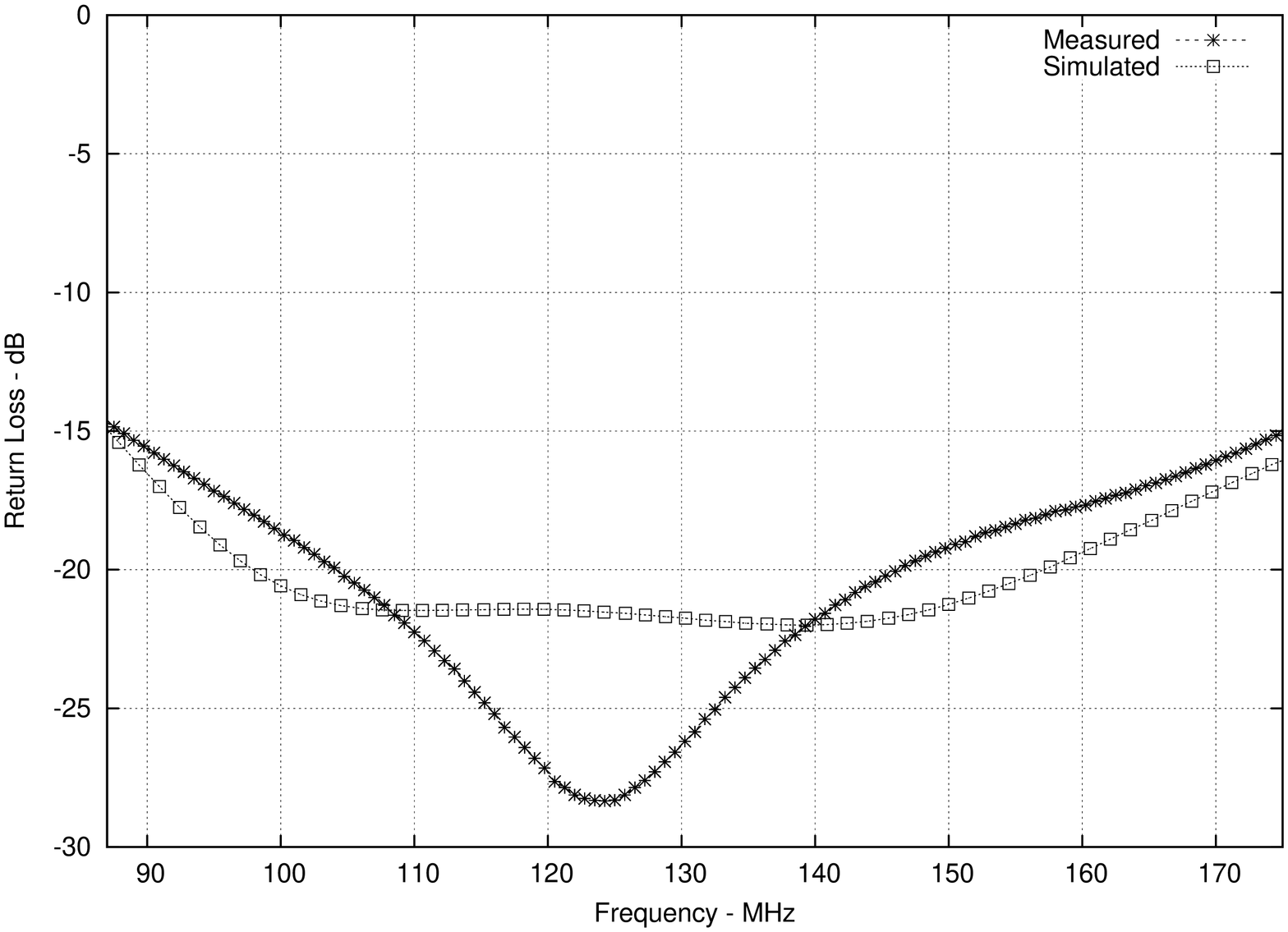}
\includegraphics[width=6.2cm,angle=-90]{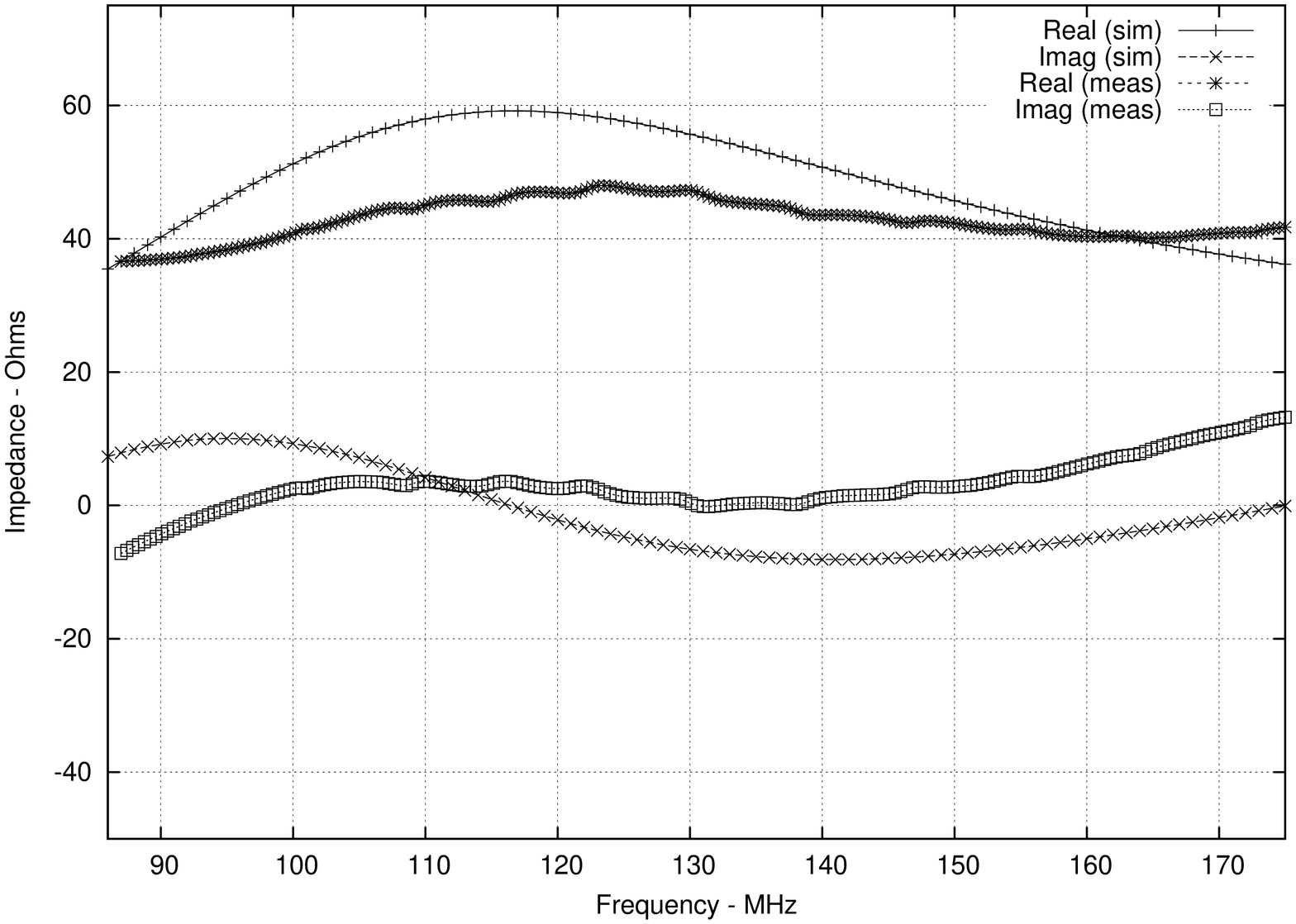}
\caption{a)~(top)~Return loss plot of antenna: Curve with square symbol represents simulation whereas the other one is the measurement. The plot indicates that the measured return loss of the antenna closely follows the
simulation over the most of the band except at the centre. It exhibits a wide band resonance around 125 MHz
~b)~(bottom)~ Real and imaginary parts of the antenna impedance: Curve with plus symbol indicates the real part of the antenna impedance obtained in simulation whereas the curve with cross symbol indicates the imaginary part. In a  similar way, the real part of the antenna impedance measured is indicated by the curve with star symbol and the imaginary part by the curve with square symbol.}}
\label{ant_RL_Re_Im}
\end{figure}

Fig.~9(a) shows the return loss plot of antenna in the frequency range 87.5--175~MHz. Curve with square symbol represents simulation while the other one is the measurement.
 The return loss measured is greater than 15 dB over the entire frequency range.
 Generally a half wave dipole tuned to a particular frequency
will have a large reactive and real impedance on either side of the resonant frequency. A thin half wave dipole 
resonant at 130 MHz will have reactive impedance ranging from $-270$~ohms 
to 375~ohms and a real part varying from 52 ohms to 246~ohms across the octave band. 
This would result in a poor impedance match with a 50~ohm transmission line. 
The fat dipole that we have designed and constructed shows superior performance both in terms of its 
real and reactive impedances.  Over the frequency range of operation, 
the real component is in the range 40 to 60~ohms and the reactive component 
is within the range  -10 to +10~ohms. This narrow impedance variation over the frequency range gives the fat-dipole 
an impedance that is fairly well matched to a 50~ohm transmission line,  resulting in a broadband performance. 
Substantial reduction in the reactive component is attributed mainly to the design profile of the fat-dipole antenna.
Fig.~9b shows the real and imaginary parts of the antenna impedance. Measurements follow the 
expectations based on the EM simulations within about $\pm$10~ohms.

\section{Measurement of currents on the antenna surface}

Our EM simulations show that the distribution of current on the antenna surface is frequency dependent.
 We measured the current distribution on the surface of the prototype using a
near-field loop antenna probe from ETS-Lindren (model No.7405).  In the experimental setup 
a CW signal derived from a standard signal generator is given to the antenna feed point.
The near field probe is scanned along the surface of  the antenna parallel to its axis. The plane of the loop probe is oriented
parallel the antenna axis in order to maximize the voltage induced by the antenna surface current.  The probe is
integrated with an amplifier so that the signal strength in the cable  at the amplifier output dominates any spurious radiative 
coupling from the antenna directly to the cable.  The amplified signal power from the probe is measured using a
spectrum analyzer.  Measurements of surface current were carried out at five frequencies across the operating band. 
 
Fig.~\ref{current_meas} shows the log-log plot of the measured current as a function of distance from the feed point. The distance in the plot is expressed in 
terms of wavelength and the vertical scale is adjusted to correct for gain at different frequencies. 
The constant slope in the plot indicates that the surface current decays in a frequency dependent way as required
for the frequency independent behaviour of the antenna.~Edge effects are thought to be responsible for the spread in the current distribution 
at the end points. The expectation based on EM simulation is shown in Fig.~\ref{current_simu}. The measurements made match 
closely the simulation results.

\begin{figure}
\includegraphics[width=6.2cm,angle=-90]{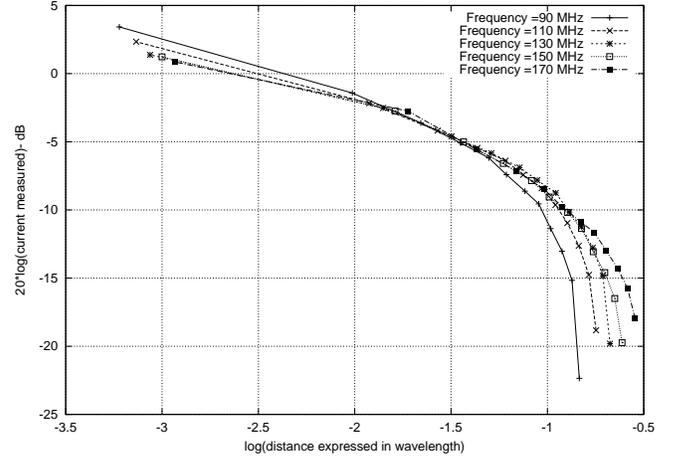}
\caption{Log-log plot showing the normalized current measured by the magnetic probe when placed on the surface at different distances from the feed point of the antenna.}
\label{current_meas}
\end{figure}

\begin{figure}
\includegraphics[width=6.2cm,angle=-90]{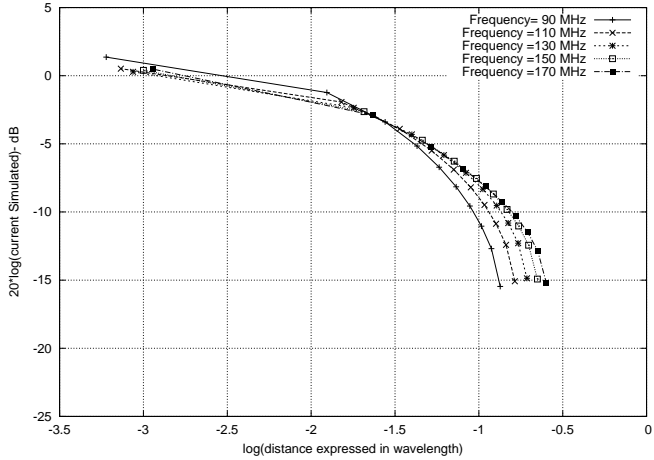}
\caption{Log-log plot showing the corresponding simulation result for the current distribution on the antenna surface.}
\label{current_simu}
\end{figure}

\section{Measurement of the radiation Pattern of the antenna}

The antenna radiation pattern was measured at five frequencies 90, 110,130,150 and 170~MHz across the
87.5 to 175~MHz band. In general, measurement of the radiation
pattern of non-directive antennas at low frequencies is challenging since
multiple reflections from nearby objects and particularly the ground poses
formidable challenges in accurately measuring the antenna
pattern. Since antennas at low frequencies tend to be bigger in size
and have larger regions of reactive fields, the zone of avoidance around
the antenna under test tends to be larger at lower frequencies
exacerbating the problem. The electrical size or cross-section of stray
wires and other metallic objects in the environment is greater at lower
frequencies, making the requirement of having a clean and large volume for
measurements important.

In the experimental set up for the measurement, the antenna under test was
used as a receiver and a short dipole antenna whose length was approximately 1/10th of the wavelength at
130~MHz, was used as the transmitter.
Both the antennas were kept off the ground and the dipole arms were oriented
parallel to the ground. The center of the fat dipole was raised  1.2~m above the 
ground.  The ground below the fat dipole and between the transmitter and receiving antennas 
were covered by ferrite absorber tiles to attenuate ground reflections. 
With the absorbers reflecting only three parts in 1000 of the incident
wave in the frequency range 87.5--175~MHz, the pattern measurement is expected to 
be accurate to  better than  $1.0\%$.  A stable signal generator model Agilent E 8257D
is used to feed the transmitter with a CW signal at discrete frequencies across
the band and at a constant power level. The receiver antenna signal is connected
to a spectrum analyzer through a co-axial cable for the power measurement.  The co-axial cable was routed below the absorber so that the measurement was not affected by the signal flowing through it. While conducting the pattern measurement, the
transmitter is kept fixed and the receiving antenna, which is the antenna under test, is rotated in azimuth over a range exceeding $180\deg$.  While making measurement, the antenna under test is mounted on a  rotating platform which had an angular scale marked on it from 0~deg. to 360~deg. in steps of 10~deg.

At all of the frequencies, the measured patterns
were symmetrical and had no significant offsets in peak response. The nulls along the dipole arms were measured to be as
deep as 27~dB below the central maximum.  The patterns over the octave band of
frequencies were very similar indicating the broad band performance of the
antenna. Figs.~\ref{pattern_dB} and \ref{pattern_lin} show plots of the radiation pattern in both logarithmic and linear units: log scale plots are
useful to show the stop band performance and depth of the nulls, linear
plots are useful in examining the pass band performance and shape of
the main lobe. The mean half power beam width (HPBW) of the measured patterns is $89.5\deg$, which matches roughly the
WIPL-D simulation result of  $85.5\deg$ (see Fig.~\ref{pattern_simu}). The deviation of the measured beamwidth from the simulation is approximately $\pm$2.3$\%$. The reason for the deviation observed is attributed partly to the measurement error and the rest to the inaccuracies in the structural dimension occurred  in the process of fabrication. 

 The change in HPBW
is measured to be $\pm$2.1$\%$ across the operating frequency range; the expectation for this 
variation based on the EM simulation is $\pm$1.4$\%$ (see Fig.~\ref{HPBW_dev}). 
Since the antenna is designed to be a short dipole, its radiation pattern is expected 
to be cosine square function.  Fig.~\ref{pattern_dev} shows the deviation in the
measured radiation pattern from this form.  The measured deviation from a cosine square pattern is 
much less than $3\%$ over the octave operating frequency range and within $\pm$
$45\deg$ of maximum response.

\section{Measurement of the absolute gain of the antenna} 
 The absolute gain of the fat dipole was measured by adopting a standard two antenna method in which a half wave dipole was used as the transmitter and the fat dipole as a receiver. While conducting the experiment, the fat dipole was kept above the absorbing tile at a distance of about 1 meter. The halfwave dipole was kept vertically above it, at a distance, more than its far field distance. The half wave dipole was connected to a standard signal generator for transmitting the signal of its resonant frequency whereas the fat dipole was connected to a spectrum analyser to measure the power received by it. Friis transmission equation was used to relate the power transmitted by the half wave dipole to the power received and estimate the gain of the fat dipole using the gain of the half wave dipole. 

The gain of the half wave dipole was independently determined using a similar approach as mentioned above in which two identical half wave dipoles were used to transmit and receive signals. The gain of the half wave dipole estimated was close to the theoretical value of 2.15 dB.  This enhanced our confidence in the estimation of the absolute gain of the fat dipole.  Since each half wave dipole designed was of narrow bandwidth in nature, different dipoles were designed to operate at 93~MHz,~122~MHz,~143~MHz and 169~MHz to estimate the absolute gain of the fat dipole over the frequency range (87.5-175)~MHz. Around each of the above frequencies, gain was estimated over 5~MHz bandwidth. The mean value of the absolute gain obtained and the standard error of the mean is shown in the  
Fig.~\ref{gain_abs} as a function of frequency along with the simulation result. The estimation of gain has been carried out after taking into account both the mismatch loss and resistive loss present in the measurement setup. The deviation of the estimated gain from the simulation result may be because of the error in the estimation of loss in the feeding section and inherent loss in the antenna structure itself.

\begin{figure}
\includegraphics[width=6.2cm,angle=-90]{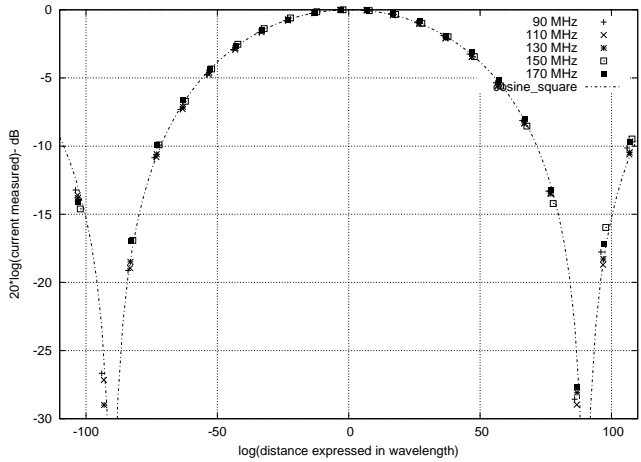}
\caption{ Radiation pattern of the wide-band fat dipole shown in logarithmic scale. Dotted line 
shows the Cosine square pattern of a short dipole for reference; symbols show
measurements at 90,110,130,150 and 170~MHz.}
\label{pattern_dB}
\end{figure}

\begin{figure}
\includegraphics[width=6.2cm,angle=-90]{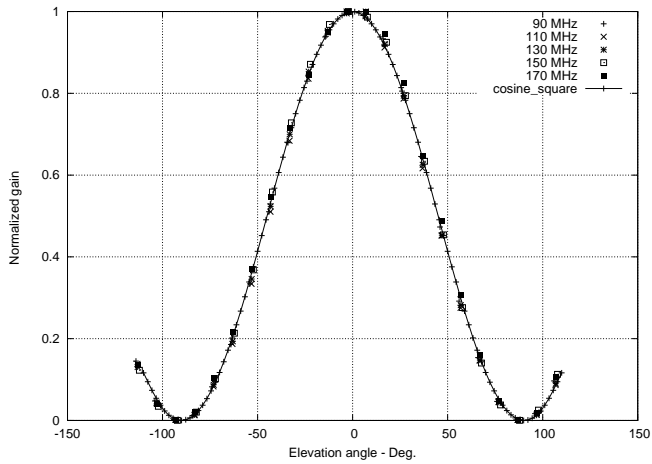}
\caption{Radiation pattern of the fat dipole shown in linear scale. Continuous line 
shows the Cosine square pattern of a short dipole for reference; symbols show
measurements at 90,110,130,150 and 170~MHz.}
\label{pattern_lin}
\end{figure}

\begin{figure}
\includegraphics[width=6.2cm,angle=-90]{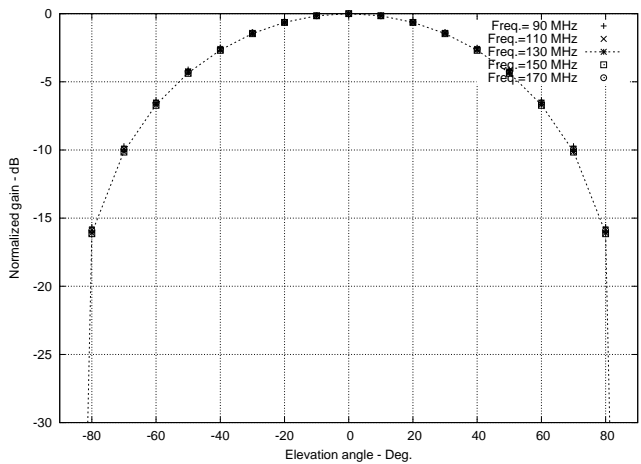}
\caption{ Simulated response of the radiation pattern of the fat dipole shown in logarithmic scale. Dotted line shows the Cosine
square pattern of a short dipole for reference; symbols show
simulation response at 90,110,130,150 and 170~MHz.}
\label{pattern_simu}
\end{figure}

\begin{figure}
\includegraphics[width=6.2cm,angle=-90]{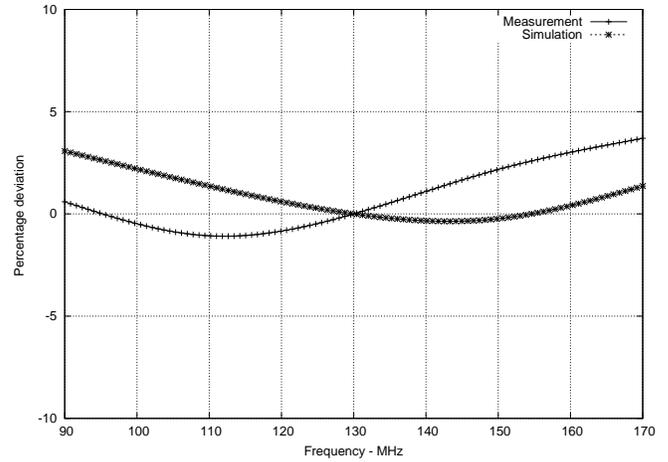}
\caption{ Antenna Half Power Beam Width variation around band centre frequency of 130 MHz. curve with plus symbol indicates
the measurement result while the other one is the simulation.}
\label{HPBW_dev}
\end{figure}

\begin{figure}
\includegraphics[width=6.2cm,angle=-90]{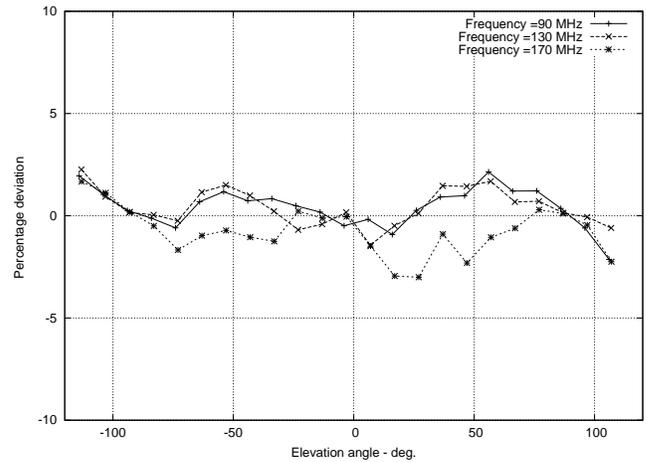}
\caption{Deviation of the measured radiation pattern from a Cosine square pattern.}
\label{pattern_dev}
\end{figure}

\begin{figure}
\includegraphics[width=6.2cm,angle=-90]{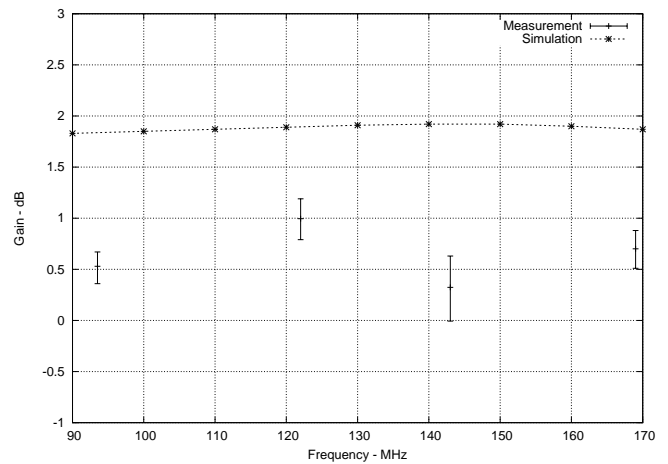}
\caption{Figure showing the absolute gain of the fat dipole as a function of frequency. Dotted line represens the simulation and the measurement results are shown by four data points with error bars.}
\label{gain_abs}
\end{figure}

\section{Summary}

We have designed and developed a low loss frequency independent fat dipole antenna having an octave bandwidth in the  frequency
range 87.5 to 175~MHz. It has a radiation pattern that varies $\pm 2.1\%$ across the frequency band 
and a return loss exceeding 15~dB.  The frequency independent performance achieved in the
impedance and radiation characteristics is attributed mainly to the sinusoidal profile given to the arms of the fat dipole.
The radiation pattern is close to a Cosine-square form and the measurement results match the results of EM simulations. The fat-dipole is a useful antenna for the measurement of the 
cosmic radio background spectrum and, in particular,  the all-sky signal from the cosmological epoch of re-ionization.

\section*{Acknowledgment}

The authors thank the members of radio astronomy laboratory of the Raman Research Institute for assistance
in the  design of balun and interconnecting cables. The mechanical engineering workshop of the Institute headed by
Ateequlla, Durai and Damodaran fabricated the antenna structure along with a styrofoam base. 
Ashwathappa and members of the Gauribidanur Observatory helped set up the field measurements.

%

\begin{biography}
[{\includegraphics[width=1in,height=1.25in,clip,keepaspectratio]{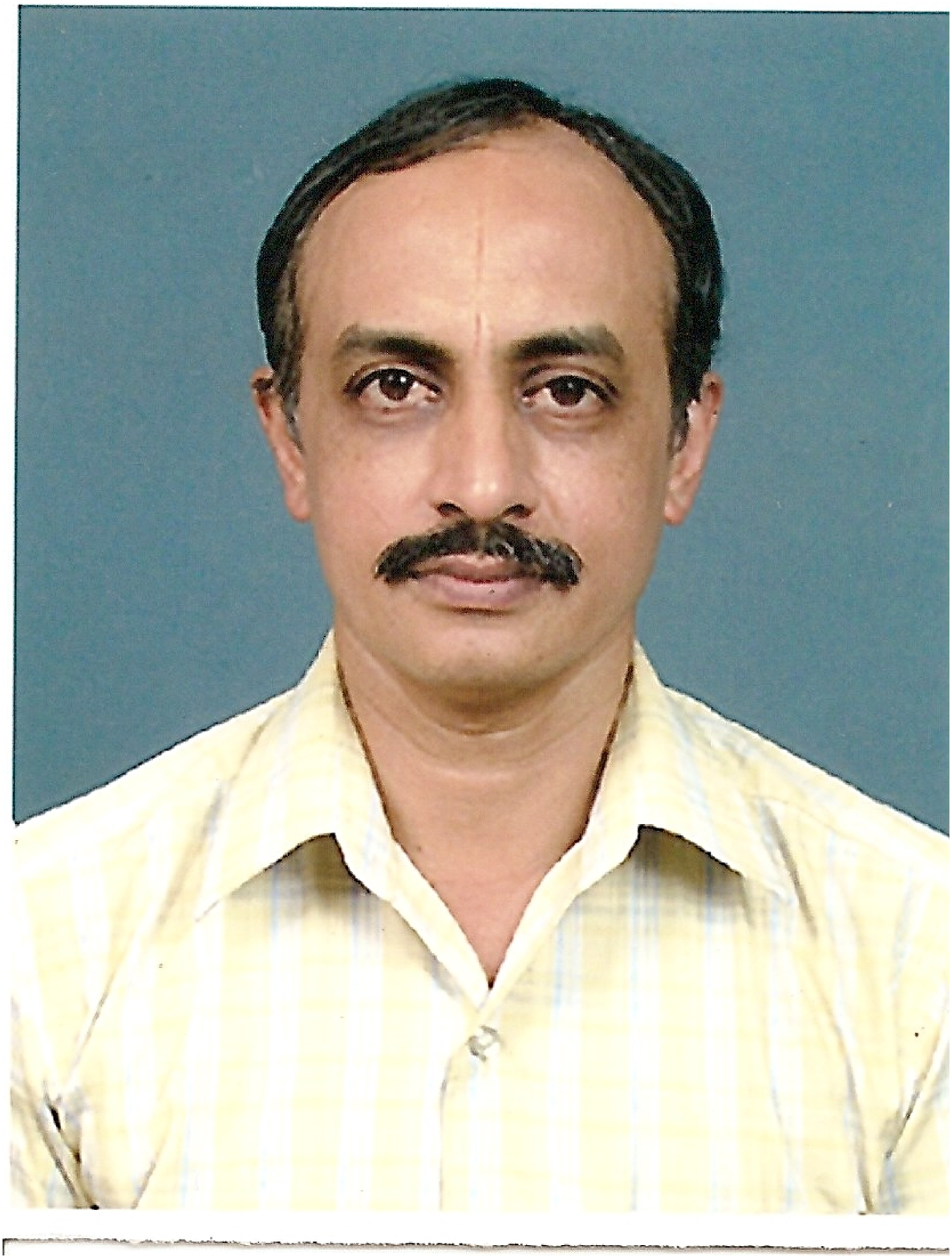}}]
{Agaram Raghunathan}
received the B.E degree in Instrumentation Engineering from Bangalore Institute of Technology,
Bangalore,1990 and M.Sc.Engg.(By Research) from Engineering dept. of the Bangalore
University, India, 2000.Since 1990,he has been at the Raman Research 
Institute, Bangalore,~India.
\end{biography}

\begin{biography}
[{\includegraphics[width=1.2in,height=1.85in,clip,keepaspectratio]{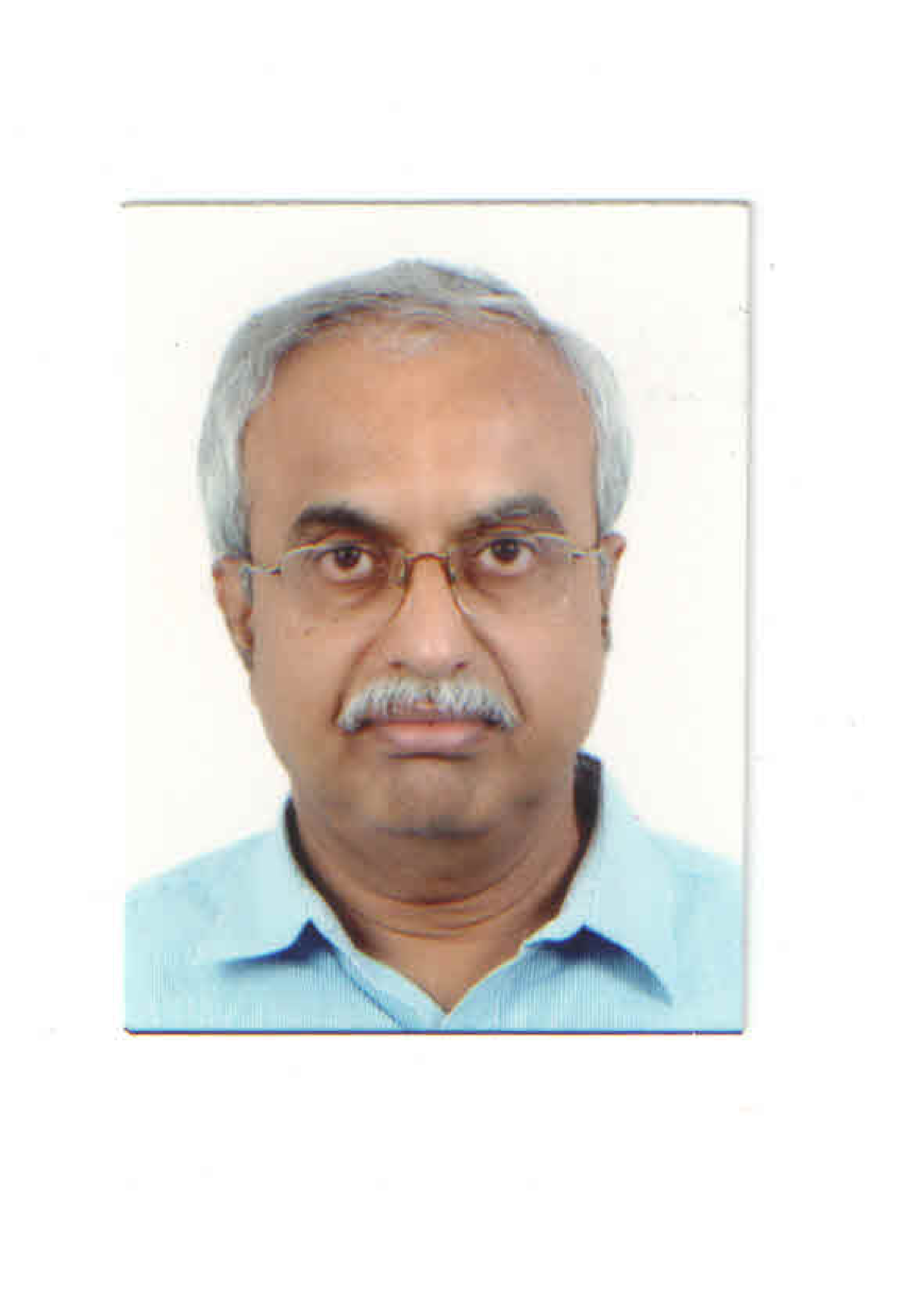}}]
{Udaya Shankar N}
 received the M.Sc. degree in Physics from the Bangalore University in 1973 and Ph.D degree in astronomy in 1986. Since 1978 he has been at the Raman Research Institute Bangalore, India, working on the instrumentation for aperture arrays, wide-field imaging and sky surveys.
\end{biography}

\begin{biography}
[{\includegraphics[width=1in,height=1.25in,clip,keepaspectratio]{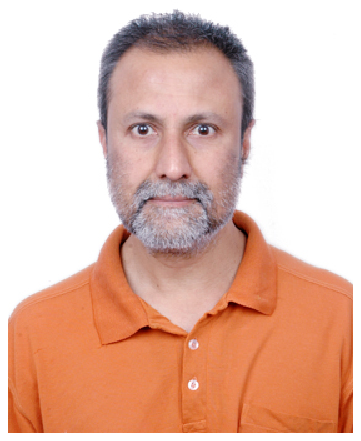}}]
{Ravi Subrahmanyan}
received the B.Tech degree in Electrical Engineering from
the Indian Institute of Technology, Madras, India, in 1983
and the Ph.D. degree in astronomy from the Physics department of the
Indian Institute of Science, Bangalore, India, in 1990.
Since 2006, he has been at the Raman Research Institute, Bangalore, India. 
\end{biography}
\vfill






\begin{thebibliography}{9}
\bibitem{}
V. H. Rumsey, "Frequency Independent Antennas," Academic press., 1966.

\bibitem{}
G. H. Brown, and O. M. Woodwarad, "Experimentally determined Radiation
characteristics of Conical and Triangular Antennas," {\it RCA Rev.},
13, pp. 425-452, Dec. 1952.

\bibitem{}
Nipanjana Patra,~Ravi Subrahmanyan,~A. Raghunathan,~N. Udaya Shankar,"SARAS: a precision system for measurement of the Cosmic Radio Background and Signatures from the Epoch of Reionization", astro-ph,arxiv:1211.3800,2012.

\bibitem{}
P. A. Shaver, R. A. Windhorst, P. Madau, and A. G. de Bruyn,"Can the
reionization epoch be detected as a global signature in the cosmic
background?," {\it Astron. Astrophys.}, vol. 345, pp. 380-390, 1999.

\bibitem{}
Pritchard, R. Jonathan, and Abraham Loeb,"Evolution of the 21cm signal
throughout cosmic history," {\it Phys. Rev. D}, vol. 78, Issue 10, id. 103511, 2008.


\bibitem{}
Adrian Liu, and  Max Tegmark, "How well can we measure and understand foregrounds
with 21cm experiment,",{\it Mon.Not.R. Astron. Soc.},vol. 419, pp. 3491-3504, 2012.


\bibitem{}
C. G. T. Haslam, C. J. Salter, H. Stoffel, and W. E. Wilson, "A 408 MHz All-Sky continuum survey. II - The aAtlas of Contour Maps,"
{\it Astron.Astrophys.Suppl.Ser.}, vol. 47, pp. 1-143, 1982.


\bibitem{}
T. L. Landecker, and R. Wielebinski, "The Galactic Metre Wave Radiation: A two-frequency survey between declinations +25$\deg$ and -25$\deg$ and 
the preparation of a map of the whole sky," {\it Aust. J. Phys. Suppl.}, vol. 16, pp. 1-30, 1970.


\bibitem{}
S. A. Schelkunoff, "Electromagnetic Waves", Van Nostrand, New York: 1943, pp. 441.

\bibitem{}
Young-Jin Park, and Jong-Hwa Song, "Development of Ultra wideband planar stepped-fat dipole antenna,"{\it Microwave and optical Technology Letters}, Vol. 48, No. 9, pp1668-1671, Sept. 2006.

\bibitem{}
Yi-Min Lu,Xue-Xia Yang, and Guo-Xin Zheng, "Analysis on a novel ultra-wide bandwidth antenna of doubleprinted circular disc," {\it Microwave and optical Technology Letters},Vol. 49, No.2, February 2007.

\bibitem{}
Joung Myoun Kim,  Young Joong Yoon, Cheol Sig Pyo, "Wideband Printed fat dipole fed by tapered microstrip balun," {\it Antennas and Propagation Society International Symposium},Vol.3,Pg. 32-35, 2003.


\end{thebibliography}
\end{document}